\documentclass[reprint,aps,prX,amsmath,amssymb,showpacs]{revtex4-1}
\usepackage{graphicx,subcaption}% Include figure files
\usepackage{dcolumn}% Align table columns on decimal point
\usepackage{bm}% bold math
\usepackage{lineno}
\usepackage{hyperref}
\usepackage{epsfig,epstopdf,float,tabularx}
\usepackage[titletoc,title]{appendix}

\begin{document}
\renewcommand{\arraystretch}{1.8}

\title{Coherent Rabi dynamics of a superradiant spin ensemble in a microwave cavity}

\author{B. C. Rose$^1$}
%\email{brose@princeton.edu}
\author{A. M. Tyryshkin$^1$}
%\email{atyryshk@princeton.edu}
\author{H. Riemann$^2$}
\author{N. V. Abrosimov$^2$}
\author{P. Becker$^3$}
\author{H.-J. Pohl$^4$}
\author{M. L. W. Thewalt$^5$}
\author{K. M. Itoh$^6$}
\author{S. A. Lyon$^1$}

\affiliation{$^1$Dept.\ of Electrical Engineering, Princeton University, Princeton, NJ 08544, USA}
\affiliation{$^2$Leibniz-Institut f{\"u}r Kristallz{\"u}chtung, 12489 Berlin, Germany}
\affiliation{$^3$PTB Braunschweig, 38116 Braunschweig, Germany}
\affiliation{$^4$VITCON Projectconsult GmbH, 07745 Jena, Germany}
\affiliation{$^5$Department of Physics, Simon Fraser University, Burnaby, British Columbia, Canada V5A 1S6}
\affiliation{$^6$School of Fundamental Science and Technology, Keio University, 3-14-1 Hiyoshi, Kohuku-ku, Yokohama 223-8522, Japan}
\date{\today}

\begin{abstract}
We achieve the strong coupling regime between an ensemble of phosphorus donor spins in a highly enriched $^{28}$Si crystal and a 3D dielectric resonator.  Spins were polarized beyond Boltzmann equilibrium using spin selective optical excitation of the no-phonon bound exciton transition resulting in $N$ = $3.6\cdot10^{13}$ unpaired spins in the ensemble.  We observed a normal mode splitting of the spin ensemble-cavity polariton resonances of 2$g\sqrt{N}$ = 580 kHz (where each spin is coupled with strength $g$) in a cavity with a quality factor of 75,000 ($\gamma \ll \kappa \approx$ 60 kHz where $\gamma$ and $\kappa$ are the spin dephasing and cavity loss rates, respectively).  The spin ensemble has a long dephasing time (T$_2^*$ = 9 $\mu$s) providing a wide window for viewing the dynamics of the coupled spin ensemble-cavity system.  The free induction decay shows up to a dozen collapses and revivals revealing a coherent exchange of excitations between the superradiant state of the spin ensemble and the cavity at the rate $g\sqrt{N}$.  The ensemble is found to evolve as a single large pseudospin according to the Tavis-Cummings model due to minimal inhomogeneous broadening and uniform spin-cavity coupling.  We demonstrate independent control of the total spin and the initial Z-projection of the psuedospin using optical excitation and microwave manipulation respectively.  We vary the microwave excitation power to rotate the pseudospin on the Bloch sphere and observe a long delay in the onset of the superradiant emission as the pseudospin approaches full inversion.  This delay is accompanied by an abrupt $\pi$ phase shift in the peusdospin microwave emission. The scaling of this delay with the initial angle and the sudden phase shift are explained by the Tavis-Cummings model.
\end{abstract}
%  By varying the optical excitation and the power of the microwave pulse we can prepare an arbitray initial state of the pseudospin in the superradiant subspace.  Not arbitray... and in that case we need to define M and S ... NO SPACE!
% To validate this model we vary the polarization of the ensemble (S=$N$/2) and find that the period of energy exchange scales as expected ($g\sqrt{N}$).  

 %change these to proper PACS

\maketitle

\begin{figure}[h]
	\includegraphics[width=\linewidth]{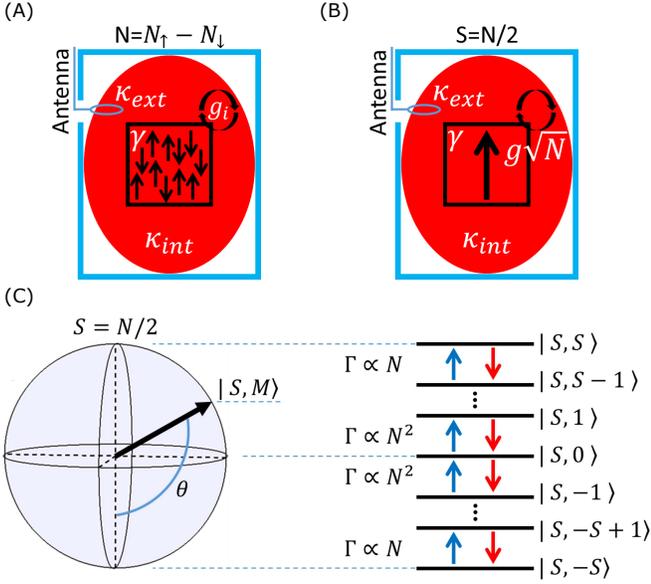}
	\captionsetup{justification=raggedright, singlelinecheck=false}
\caption{(A)~Spin ensemble with individual spin-cavity coupling $g_i$ ($i$=1,...,$N$).  Dissipation ($\kappa_{int}$) and coupling losses ($\kappa_{ext}$) in the cavity along with dephasing ($\gamma \propto 1/T_2^*$) in the spin ensemble set the maximum window during which the spin-cavity dynamics can be probed. (B)~Large pseudospin model (Dicke model) assuming small inhomogeneous broadening ($\omega_i=\omega$) in the ensemble and uniform cavity coupling ($g_i$=$g$). (C)~Bloch sphere representation of the pseudospin.  The sphere surface represents the symmetric superradiant subspace ($\mid S,M\rangle$).  Strong correlations between spins in these states lead to an enhanced photon emission rate $\Gamma\propto$ $N^2$ near the equator ($M=0$).  Uncorrelated spontaneous emission at a rate $\Gamma\propto N$ dominates near the poles ($M = \pm S$).}
\label{fig:TCCartoon} 
\end{figure}

%\section{\label{sec:Intro}Introduction}
% In quantum computing, strong coupling between the spin-ensemble and cavity is necessary for coherent exchange of quantum information.\cite{Ritsch2011, Kubo2010, Schuster2010,2013Probst} 
The enhanced collective emission from an ensemble of atom-like systems due to coherent self-stimulated emission was originally described by Dicke\cite{Dicke1954}.  For this phenomenon he coined the term superradiance and showed that the atom ensemble can behave as a large collective pseudospin.  Superradiant emission has been observed in numerous physical systems.\cite{Gross1982301,Haroche1985347, Skribanowitz1973, Gross1976, Scheibner2007, Chalupczak2015}  These collective effects are particularily prominent when the coupling between the spin-ensemble and the radiation field ($g\sqrt{N}$ for $N$ spins individually coupled with strength $g$) is larger than any of the losses in the system ($\kappa + \gamma$ where $\kappa$ is the radiative loss rate and $\gamma$ is the spin dephasing rate).  This strong coupling regime has been extensively studied in both theory and experiment,\cite{Zhu1990,Hettich385, Baumann2010, Diniz2011} but clearly resolved dynamics of these collective effects are generally lacking in large ensembles due to strong dephasing (short T$_2^*=2\pi/\gamma$).\cite{1997Mielke,1996BruneHaroche, Gross1982301}  Here we demonstrate the dynamics of a strongly coupled ensemble of phosphorus donor spins in highly isotopically enriched $^{28}$Si with both a long dephasing time\cite{Tyryshkin2012,2010Becker} and uniform coupling to the radiation field (due to the use of a 3D microwave cavity) as shown schematically in Fig. \ref{fig:TCCartoon}A.  For the first time outside of ensembles of Rydberg atoms\cite{1996BruneHaroche, 1997Mielke}, we have studied a spin ensemble-cavity system with both of these essential properties allowing for it to be modeled accurately as a single large pseudospin (Fig. \ref{fig:TCCartoon}B) offering a simple interpretation of the spin ensemble-cavity evolution.\cite{Dicke1954,TavisCummings1968}  In particular, we are able to directly observe the dynamics of superradiant emission\cite{1983KaluznyHaroche} under strong excitation.  In contrast to the more extensively studied low excitation limit, with strong excitation the system cannot be modeled as a linear system of two coupled harmonic oscillators.\cite{1993Thompson, 1990Zhu}  Instead it must be treated with the Tavis-Cummings model.\cite{TavisCummings1968} 

	Initialization of the pseudospin was accomplished with a combination of optical excitation which sets the size of the pseudospin, $S=N/2$, and microwave manipulation which sets the initial z-projection of the pseudospin, $M$.  Previous reports of large superradiant ensembles only prepared an initially inverted state $\mid S, S\rangle$,\cite{1997Mielke,1996BruneHaroche} while with our system we are able to prepare any initial pseudospin state $\mid S, M\rangle$ in the superradiant subspace (Fig. \ref{fig:TCCartoon}C) with independent control of S and M.  By controlling the size of the pseudospin we demonstrate a $g\sqrt{N}$ dependence of the energy exchange rate between the pseudospin and cavity.  By varying M, we are able to control a delay in the onset of the superradiant emission\cite{1981Haake} and we report the first experimental observation of a log dependence of this delay when the pseudospin is near quasi-equilibrium at $M=S$.  This log dependence is consistent with predictions from the Tavis-Cummings model.  We also observe an abrupt $\pi$ phase shift in the pseudospin microwave emission around this fully inverted state $M=S$. 

\begin{figure}[h]%NEED h and NOT H ...
\centering
\includegraphics[width=0.88\linewidth]{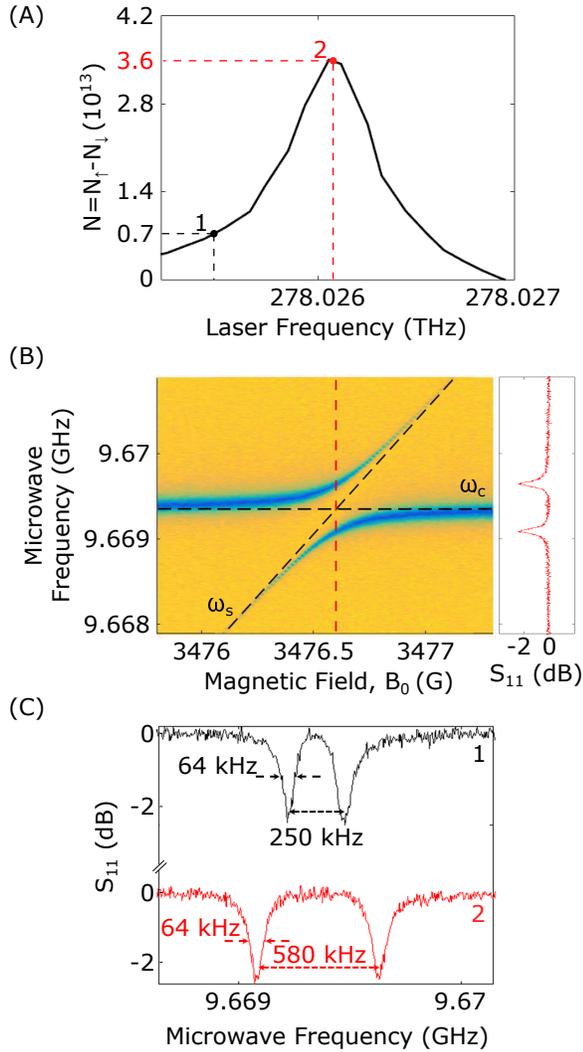}
\captionsetup{justification=raggedright,singlelinecheck=false}
\caption{(A)~The number of spins interacting with the resonator ($N$) is controlled by tuning the laser frequency which is spin-selectively exciting a phosphorus donor no-phonon bound exciton transition.\cite{2009YangThewalt} (B)~Reflected microwave power $S_{11}$ (color scale) spectrum of the cavity as a function of the spin ensemble Zeeman splitting ($\omega_s=g\mu_B B_0$) as the spin ensemble is brought into resonance with the bare cavity ($\omega_c$).  The clear avoided crossing shows that the system is in the strong coupling regime with $2g\sqrt{N}$ (580 kHz) $\gg\kappa+\gamma$ (64 kHz). (C)~The vacuum Rabi splitting ($2g\sqrt{N}$) of the polariton modes (shown at $\omega_s=\omega_c$) can be controlled by detuning the laser frequency.  The curves 1 and 2 were measured while the laser frequency was tuned to points 1 and 2 as indicated in panel (A).}
\label{fig:StationaryStates} 
\end{figure}

\begin{figure}[h]
\includegraphics[width=0.9\linewidth]{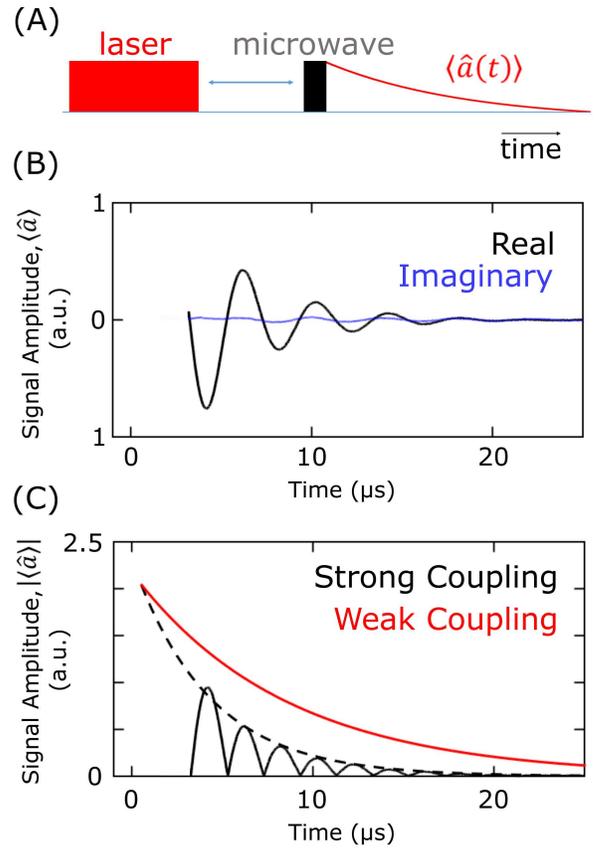}
\captionsetup{justification=raggedright,singlelinecheck=false}
\caption{(A)~Pulse sequence of the free induction decay (FID) experiment.  The initial laser pulse sets the number of unpaired spins $N$ and the total spin quantum number $S = N/2$, the subsequent microwave pulse excites the pseudospin to $\mid S=N/2, M\rangle$.  The microwave emission from the pseudospin is measured after the microwave pulse. (B)~In-phase, Re$\left(\hat{a}\right)$, and quadrature, Im$\left(\hat{a}\right)$, components of the pseudospin microwave emission during the free induction decay in the strong coupling regime ($M \approx 0$, $N = 3.6\cdot10^{13}$). (C)~Amplitude of the microwave emission, $\mid \hat{a}\mid$, during the FID.  In the weak coupling regime the decay is purely exponential (red line).  In the strong coupling regime the FID shows oscillations at rate $g\sqrt{N}$ resulting from the vacuum Rabi splitting between the two polariton modes.  These oscillations occur as excitations are exchanged between the spin ensemble and the cavity.  An experimental dead time after the microwave pulse of 3 $\mu$s limits us from measuring the beginning the FID.
}
\label{fig:FIDTimeExp} 
\end{figure}

	The isotopically enriched silicon crystal ($<$50 ppm~$^{29}$Si) used in these experiments is phosphorus doped with a density of 3.3$\cdot$10$^{15}$ cm$^{-3}$ (5.7$\cdot$10$^{13}$ total donors) and is otherwise highly pure (boron density less than 10$^{14}$ cm$^{-3}$).\cite{2010Becker}  A tunable distributed Bragg reflector (DBR) laser (Eagleyard EYP-DBR-1080) was used to controllably polarize the phosphorus donor spin ensemble beyond Boltzmann equilibrium by spin-selective optical pumping of the phosphorus donor no-phonon bound exciton transitions.\cite{2009YangThewalt}  The efficiency of the optical pumping and the resulting steady state spin polarization ($N = N_\uparrow-N_\downarrow$) was controlled by detuning the laser from resonance with one of the no-phonon bound exciton transitions (Fig. \ref{fig:StationaryStates}A).  The laser was tunable between 1077-1081nm (277.2-278.5 THz) and fiber coupled to deliver $\sim$10~mW of light to the sample.  More than 95$\%$ electron spin polarization was achieved after 300 ms of resonant optical pumping (Fig. \ref{fig:StationaryStates}A).  The $^{31}$P nuclei were also polarized to $\approx 25\%$ during the optical pumping.  The mechanism behind this nuclear polarization remains unknown but it is thought to be due to an enhanced cross relaxation rate of the donors under illumination.\cite{Gumann2014}  X-band (9.6~GHz) ESR experiments were performed with a Bruker ESR spectrometer (Elexsys E580) using a dielectric 3D resonator (ER-4118X-MD5) in a helium-flow cryostat (Oxford CF935).  In the free induction decay experiments a single 200 ns microwave pulse was used to tip the spins after polarizing with the DBR laser (Fig. \ref{fig:FIDTimeExp}A).  The tipping angle of the microwave pulse was controlled by varying the power of the microwave source (Agilent E8267D).

	The stationary states of the system were interrogated by measuring reflected microwave power ($S_{11}$) with an Agilent E5071C network analyzer (Figs. \ref{fig:StationaryStates}B-C).  Power saturation was avoided by lowering the probe power to -65 dBm (power incident on the resonator); below this power the measured vacuum Rabi splitting was constant and no power broadening was observed.  The $S_{11}$ measurements were performed under continuous illumination from the resonant DBR laser which in addition to polarizing the spin ensemble accelerates spin relaxation (T$_1$) and reduces power saturation from the microwave probe.

	The total ensemble coupling can be determined directly by looking at the eigenfrequencies of the coupled system while the spin ensemble is tuned into resonance with the cavity (Fig. \ref{fig:StationaryStates}B).  This technique has been demonstrated for spin ensembles coupled to both 3D volume resonators \cite{2011Eisuke} and 2D superconducting microresonators\cite{2010Schuster}.  The spin transition frequency ($\omega_s$) is varied through resonance with the cavity ($\omega_c$) by changing the Zeeman splitting with a magnetic field (B$_0$) which resulted in an avoided crossing with a clear splitting showing that the system is in the strong coupling regime.  In particular, with the laser tuned on resonance ($N$ = 3.6$\cdot10^{13}$) the vacuum Rabi splitting of the polariton modes is 580 kHz ($2g\sqrt{N}$) which is an order of magnitude larger than their 64 kHz linewidth (Fig. \ref{fig:StationaryStates}C, curve 2).  The splitting can be controllably reduced by detuning the laser from resonance with the bound exciton transition (curve 1 in Fig. \ref{fig:StationaryStates}C).  The large size of the ensemble is important to achieve this large splitting since each spin is weakly coupled to the 3D cavity with a single spin coupling $g$ = 38 mHz.  The narrow linewidth of the resonances is defined by combined losses in the cavity ($\kappa$ = 60 kHz for Q = 75,000) and dephasing in the spin ensemble ($\gamma$ = 18 kHz).  The ensemble is in the high cooperativity limit ($C=g^2N/\kappa\gamma \approx 280$) where energy can be exchanged efficiently between the spins and the cavity.  
%This ratio is largely due to the use of highly enriched $^{28}$Si ($<$50 ppm $^{29}$Si) which is the state of the art in isotopic enrichment originating from efforts to develop a new mass standard.\cite{2010AvogadroProject} 

	The free evolution of the pseudospin in a cavity is most directly studied in a single pulse free induction decay (FID) experiment (Fig. \ref{fig:FIDTimeExp}A).  The pseudospin is first polarized into its ground state ($M=-S$) with resonant laser pumping thus defining $N=N_\uparrow-N_\downarrow$ and $S=N/2$.  This is followed by a single microwave pulse that determines the initial state (M) of the pseudospin before the FID.  In the weak coupling regime the envelope of the free induction decay is a simple exponential (Fig. \ref{fig:FIDTimeExp}A) with a characteristic decay time resulting from the inhomogeneous broadening of the spin ensemble (T$_2^*$).  However, in the strong coupling limit achieved here (Fig. \ref{fig:FIDTimeExp}B-C) the free induction decay shows multiple oscillations as energy is coherently exchanged between the spins and the resonator through their coupling at a rate of $2g\sqrt{N}=$ 580 kHz corresponding to the vacuum Rabi splitting of the polariton modes.  

	In the strong coupling regime, the envelope decay of the polariton modes (Fig. \ref{fig:FIDTimeExp}C, dashed black line) is faster than the decay in the weak coupling regime as it results from a combination of spin dephasing plus resonator losses as the excitations are exchanged between the spin ensemble and the cavity over time.  Spin relaxation (T$_1$) of phosphorus donors at 1.5 K is slow and does not enter into the dynamics of the free induction decay.\cite{Tyryshkin2012}  Even with this faster decay we are able to observe 12 collapses and revivals before the free induction decay falls below the noise level.  An experimental dead time of 3 $\mu$s after the microwave pulse limits us from measuring the beginning of the FID signal.  The large signal to noise and many oscillations observed here offer clearly resolved dynamics of the coupled spin-ensemble/cavity evolution.

	A theoretical description of this experiment begins with a general model for an ensemble of $N$ spin 1/2 particles interacting with a single cavity mode:
\begin{equation}
\begin{split}
\hat{H}=\sum\limits_i \omega_i \hat{s}_{Zi} &+ \omega_c \hat{a}^\dagger \hat{a}+2\sum\limits_i g_i(\hat{a}^\dagger+\hat{a})\hat{s}_{Xi}\\
&+\sum\limits_{<ij>}\bf{\hat{s}}_i\cdot \bf{\hat{D}}_{ij}\cdot\bf{\hat{s}}_j
\end{split}
\label{eq:Dicke}
\end{equation}
%Drive term +(\eta(t)\hat{a}^\dagger e^{-i\omega_d t} +\eta^*(t) \hat{a} e^{i\omega_d t})
where $\hat{a}^\dagger$, $\hat{a}$ are the creation and annihilation operators for the cavity field photons of frequency $\omega_c$, $s_{ki}$ ($i$ = 1,...,$N$ and $k$ = X,Y,Z) are the single spin ($S = 1/2$) matrices for a spin with transition frequency $\omega_i$ and spin-cavity coupling $g_i$.  The last term ($\bf{\hat{D}}_{ij}$) describes dipolar interactions between spins.  

\begin{figure}[h]
\includegraphics[width=0.9\linewidth]{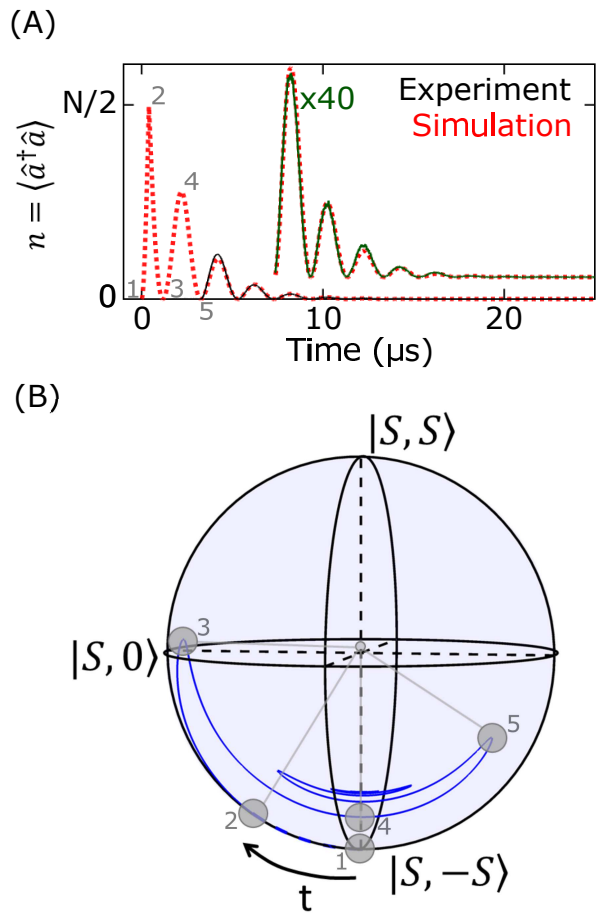}
\captionsetup{justification=raggedright,singlelinecheck=false}
\caption{Collective dynamics between the pseudospin and cavity. (A) Cavity photon number ($n(t)=\langle\hat{a}^\dagger\hat{a}\rangle$) during the FID experiment (the square of the FID amplitude shown in Fig. \ref{fig:FIDTimeExp}C) with $N = 3.6\cdot 10^{13}$ and $M \approx 0$.  The simulated fit according to Eq. \ref{eq:HeisenExpectation} is also shown (dashed curve).  The tail of the decay is magnified x40 and shown in green. (B)~The pseudospin evolution on the Bloch sphere as determined from $\langle S_-(t) \rangle$ and  $\langle S_z(t) \rangle$ in this simulation.  The numbers (1-4) on the curves indicate equivalent points in time during the FID on the microwave emission plot (A) and the simulated pseudospin evolution plot (B).  The dynamics of the pseudospin in this large $N$ limit is governed by the Maxwell-Bloch equations.
}
\label{fig:FIDTimeSim} 
\end{figure}

	Direct diagonalization of this Hamiltonian for large spin ensembles is not possible, however several simplifying approximations can be made to good accuracy.  The spin ensemble has a narrow inhomogeneous distribution in Zeeman frequencies compared to the spin-cavity coupling ($g\sqrt{N}$T$_2^*$ $\gg 1$) so that, for the duration of our FID experiment, we can treat the spins as having identical transition frequencies $\omega_i\approx$ $\omega_s$.  The distribution in the individual spin-cavity coupling ($g$) is mostly defined by the microwave magnetic field inhomogeneity along the length of the sample (5 mm).  For our volume resonator this variation has been measured to be less than 5$\%$.\cite{MortonRotationErrors}  Thus to a good approximation the spin ensemble-cavity system is in the small sample limit with a uniform spin-cavity coupling ($g_i\approx g$).\cite{1982RaimondandHaroche, TavisCummings1968}  Finally, the dipolar coupling between spins (last term in Eq. \ref{eq:Dicke}) at the 3.3$\cdot$10$^{15}$ cm$^{-3}$ donor density used in this experiment is $\sim$100 Hz and negligible on the 30 $\mu$s timescale measured here.\cite{2012Witzel}  With both the Zeeman frequency and the cavity coupling being constant across all of the spins, the ensemble can be treated as a single large pseudospin with a collective spin operator $\hat{S}_{(z,\pm)}=\sum\limits_i\hat{s}_{(z,\pm)i}$.  Including all of these considerations, Eq. \ref{eq:Dicke} reduces to the Tavis-Cummings model:

\begin{equation}
\hat{H}_{TC}= \omega_s\hat{S}_{Z} + \omega_c\hat{a}^\dagger \hat{a}+g(\hat{a}^\dagger\hat{S}_{-} + \hat{a}\hat{S}_{+})
\label{eq:TC}
\end{equation}
%Drive term &+(\eta(t)\hat{a}^\dagger +\eta^*(t) \hat{a}
which is a generalization of the Jaynes-Cummings model for a single collective pseudospin $\hat{S}$.  From this model we derive the equations of motion for the pseudospin-cavity system:
\begin{equation}
\begin{split}
&\langle \dot{\hat{a}}(t)\rangle = -\kappa \langle \hat{a}(t)\rangle-ig \langle \hat{S}_-(t)\rangle-iV(t)\\
&\langle\dot{\hat{S}}_-(t)\rangle = -\gamma  \langle \hat{S}_z(t)\rangle+2ig \langle \hat{a}(t)\rangle \langle \hat{S}_z(t)\rangle\\
&\langle \dot{\hat{S}}_z(t)\rangle = i g\left( \langle a^\dagger(t)\rangle \langle S_-(t) \rangle-\langle a(t)\rangle \langle S_-(t) \rangle^* \right)
\end{split}
\label{eq:HeisenExpectation}
\end{equation}
These are the Maxwell-Bloch equations for the expectation values $\langle \hat{a}\rangle$,  $\langle S_- \rangle$, and $\langle S_z \rangle$.\cite{1965MaxwellBloch}  The dissipation factors, $\kappa$ and $\gamma$, were introduced using the standard master equation formalism for the open system.\cite{qMechDissipativeSystems}  We have also added a classical drive term with amplitude $V(t)$ set by the microwave source.  To simplify these equations we take the semiclassical limit by neglecting correlations between the spin ensemble and cavity photons  ($\langle a S_i \rangle = \langle a\rangle\langle S_i \rangle$ for $i=+,-,z$).  This approximation is valid given the large size of our spin ensemble ($N\sim10^{13}$ spins).  We use Eq. \ref{eq:HeisenExpectation} to simulate the dynamics of the free induction decay (Fig. \ref{fig:FIDTimeSim}).

	In the absence of dephasing ($\gamma=0$) the total spin of the ensemble is conserved ($\left[\hat{H}_{TC},\hat{S}^2\right]=0$) and the system evolution is a trajectory on the surface of the Bloch sphere within the superradiant subspace (Fig. \ref{fig:TCCartoon}C).  Dephasing from inhomogeneous broadening (T$_2^*$) opens up a channel for mixing with subradiant states so that $S$ is no longer preserved and the system enters states in the interior of the sphere (Fig. $\ref{fig:FIDTimeSim}$B).\cite{Temnov2005}  In our experiment this dephasing is not refocused and thus is irreversible, so that it can be taken as a single loss rate, $\gamma$, for the transverse magnetization of the spin ensemble.  The value for $\gamma$ (also used in the  $S_{11}$ measurements) was extracted directly from the free induction decay in the weak coupling regime where spin dephasing is the only contribution.  Photon losses are also taken into account with a single parameter $\kappa=\kappa_{ext}+\kappa_{int}$ which represents both internal loss in the cavity ($\kappa_{int}$) and external loss through the antenna coupling ($\kappa_{ext}$).  This parameter was extracted by measuring the linewidth of the cavity resonance under low excitation where strong coupling effects were negligible.  The 64 kHz polariton linewidth in Fig. \ref{fig:StationaryStates}C is a combination of both $\kappa$ and $\gamma$, with $\kappa\gg\gamma$.

A simulation of the coupled system according to these equations of motion is compared to the measured curve in Figure \ref{fig:FIDTimeSim}A.  Here we plot the intensity of the microwave emission (cavity photon number, $n=\langle\hat{a}^\dagger\hat{a}\rangle$) derived from the microwave amplitude which was measured directly and plotted in Fig. \ref{fig:FIDTimeExp}C.  The only fitting parameter used in this simulation was a scaling parameter for the microwave pulse ($V(t)$).  This parameter accounts for the resonator coupling as well as losses in the microwave excitation path.  The same scaling parameter was used consistently for all microwave powers set in our experiments.  The simulation shows an excellent fit to the measured FID curve (Fig. \ref{fig:FIDTimeSim}A) with the main difference between the two curves being only the experimental dead time (initial 3 $\mu$s).  The time evolution of the pseudospin state as determined from this fit is plotted in Figure \ref{fig:FIDTimeSim}B.

	The motion of the pseudospin on the Bloch sphere is formally equivalent to a damped pendulum that is kicked into motion by the microwave excitation.\cite{1983KaluznyHaroche}  The pseudospin is initially polarized to its ground state at the south pole (point 1 on the Bloch sphere in Fig. \ref{fig:FIDTimeSim}B).  A microwave pulse is applied for 200 ns (black arrow), mainly populating the cavity with photons but also starting the initial excitation of the pseudospin (point 2).  The driving microwave pulse ends but the remaining photons in the cavity continue to be transfered to the pseudospin until no photons remain (point 3 on the Bloch sphere).  The power of the applied microwave pulse in this example was chosen so that this point would be close to the equator ($\mid S, 0\rangle$).  This point (3) corresponds to a minimum ($n=0$) in the free induction decay (Fig. \ref{fig:FIDTimeSim}A, point 3) since the photons in the resonator have been fully absorbed by the pseudospin.  After reaching this highest point, the excited pseudospin will start to emit photons back into the resonator, reversing the nutation on the Bloch sphere to transition towards its ground state (point 4 in Fig. \ref{fig:FIDTimeSim}B).  During the initial excitation the pseudospin (up until point 3) evolves with the same phase as the microwave pulse, but as the pseudospin reverses direction the photons it emits have a phase opposite to that of the initial pulse.  The pseudospin eventually fully de-excites to $\mid S, -S\rangle$ (point 4), releasing all of the photons back into the resonator.  This point corresponds to a maximum in the free induction decay (Fig. \ref{fig:FIDTimeSim}A, point 4).  The pseudospin then reabsorbs these photons and is excited to the opposite side of the Bloch sphere (point 5), since the emitted photons are of opposite phase to the initial excitation.  This coherent exchange of excitations continues until dissipation and dephasing destroy the ensemble polarization.

	Near the maxima and minima of the FID, where $\dot{n}=0$, the pseudospin evolution is dominated by the second derivative (in time) of the cavity photon number ($\ddot{n}$).  In the strong coupling regime (ignoring the dissipation terms) this can be approximated as:
\begin{equation}
\ddot{n} \approx  g^2 \left( \langle \hat{S}_+\hat{S}_-\rangle + \langle \hat{S}_-\hat{S}_+\rangle\right)+4g^2\left(n + 1/2\right)\langle \hat{S}_z \rangle
\label{eq:NConcavity}
\end{equation}

With this expression we can explain the shape of the FID near each extrema.  At minima (i.e. point 3) there are no photons in the cavity ($n\approx 0$) and also $\langle \hat{S}_z\rangle=0$, therefore the second term in Eq. \ref{eq:NConcavity} is small.  However, the pseudospin is near $\mid S, 0\rangle$ where correlations between spins lead to superradiant emission with an emission rate that is quadratic in the total number of unpaired spins\cite{Dicke1954} so that the first term in Eq. \ref{eq:NConcavity} is large, $\ddot{n}\propto g^2 \langle\hat{S}_+\hat{S}_-\rangle\propto g^2N^2/2$.  At maxima (i.e. point 4) the pseudospin is near $\mid S, -S \rangle$  ($\langle \hat{S}_z\rangle\approx -N/2$) where the correlations between spins in the ensemble are negligible and $\langle\hat{S}_-\hat{S}_+\rangle\approx N$ (similar to spontaneous emission of N uncorrelated emitters).  However, most of the photons have transfered back into the cavity ($n\approx N$) so that the second term in Eq. \ref{eq:NConcavity} is now large, resulting in a similarly fast $\ddot{n}\propto -g^2N^2$.  The pseudospin is re-excited by the large population of photons in the resonator ($n \approx N$), but dissipation and dephasing eventually destroy the ensemble polarization ($N$), terminating the FID. 

	The FID we observed is different from superradiance phenomena in free space where the system is far from the strong coupling regime.  In free space the emitted photons leave the system much faster than they can be reabsorbed by the pseudospin which means that the second term in Eq. \ref{eq:NConcavity} can be neglected.  As a result, near $M=-S$ the pseudospin evolves slowly with $\ddot{n} \propto g^2 N$.  This is a factor of N smaller than the value of $\ddot{n}$ we observe in the strong coupling regime.  Additionally, in free space the pseudospin emission dies out entirely after the ground state $\mid S, -S\rangle$ is reached because there are no excitations left in the system ($n=0$) to be reabsorbed by the pseudospin.  

\begin{figure}[t]%NEED h and NOT H ...
\centering
\includegraphics[width=\linewidth]{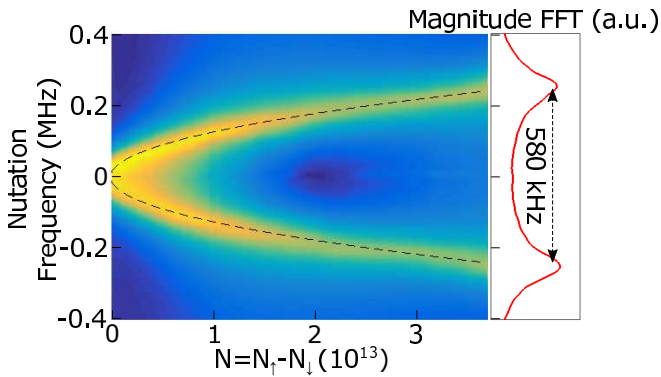}
\captionsetup{justification=raggedright,singlelinecheck=false}
\caption{Fourier transform (plotted as magnitude) of the donor spin ensemble free induction decay as a function of the number of unpaired spins in the ensemble ($N$).  The black dashed line shows a $g\sqrt{N}$ dependence of the pseudospin-cavity exchange rate as expected from the Tavis-Cummings model in the large $N$ limit.  This gives a single spin coupling of $g$=37 mHz for a uniformly coupled ensemble.  The side-plot shows a vertical slice at $N = 3.6\cdot10^{13}$ cm$^{-3}$ for the maximum ensemble coupling that was achieved ($2g\sqrt{N}$ = 580 kHz).}
\label{fig:FIDDepPol} 
\end{figure}

\begin{figure}[h]%NEED h and NOT H ...
\centering
\includegraphics[width=\linewidth]{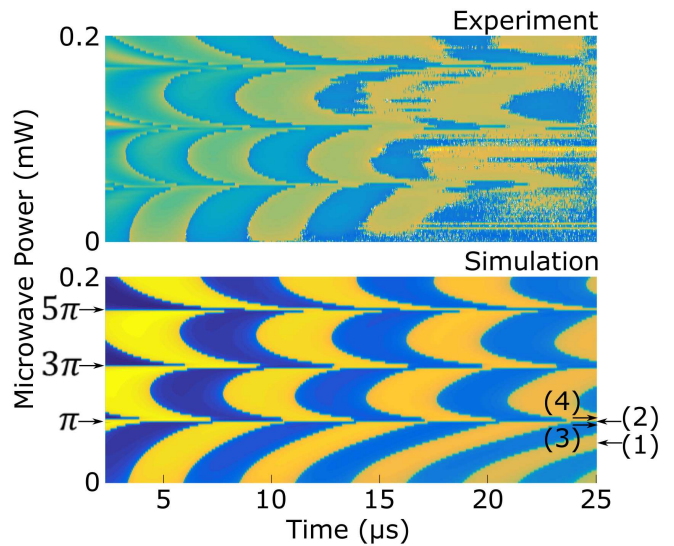}
\captionsetup{justification=raggedright,singlelinecheck=false}
\caption{(Top)~In-phase component (Re$\langle \hat{a}\rangle$) of the pseudospin microwave emission during the FID experiment (time axis) plotted as a function of the rotation angle of the initial microwave pulse (microwave power axis).  The signal intensity is plotted using a log colorscale in order to emphasize the sign of the signal.  The blue/yellow colors correspond to negative/positive signal intensities. (Bottom)~The corresponding FID simulations using the Maxwell-Bloch equations (Eq. \ref{eq:HeisenExpectation}).  The rotation angles ($\theta=\pi, 3\pi, 5\pi$) indicated on the left correspond to the inverted state of the pseudospin, $\mid S,S\rangle$.  The numbered arrows (1)-(4) shown on the right correspond to FID experiments shown separately in Fig. \ref{fig:FIDmwPowerDep1DTrace}.}
\label{fig:FIDmwPowerDep} 
\end{figure}

	In order to better understand the dynamics in this strong coupling regime we varied the initial state of the pseudospin to observe its effect on the subsequent evolution.  The initial state of the pseudospin, $\mid S, M\rangle$, can be accurately controlled both by adjusting the detuning frequency of the laser which determines the total spin $S$ (Fig. \ref{fig:StationaryStates}A), and by changing the power of the microwave pulse, which determines $M$.  The observed dependencies on $S$ and $M$ are plotted in Figs. \ref{fig:FIDDepPol} and \ref{fig:FIDmwPowerDep}, respectively. 

	Varying the laser detuning allows for fine control of the net ensemble polarization ($N$) and the total spin ensemble-cavity coupling ($g\sqrt{N}$).  The magnitude of the Fourier transform of the free induction decay (Fig. \ref{fig:FIDDepPol}) reveals that the pseudospin-cavity exchange frequency scales as $g\sqrt{N}$ (dashed black) in agreement with the large pseudospin picture.  The shapes of the polariton peaks in the Fourier transform of the free induction decay (side plot in Fig. \ref{fig:FIDDepPol}) are less resolved than the shapes obtained by reflection spectroscopy (Fig. \ref{fig:StationaryStates}C).  This is due to the fact that during the free induction decay a coherence, created by the initial microwave pulse, leaks to subradiant states ($S<N/2$) on the timescale of $T_2^*$.  This leakage gives rise to a time dependent change of $N$ during the FID and decreases the cavity coupling of the pseudospin, effectively smearing out the peaks in the Fourier transforms.  This leakage does not occur in the reflection measurements.  

\begin{figure}[h]%NEED h and NOT H ...
\centering
\includegraphics[width=0.8\linewidth]{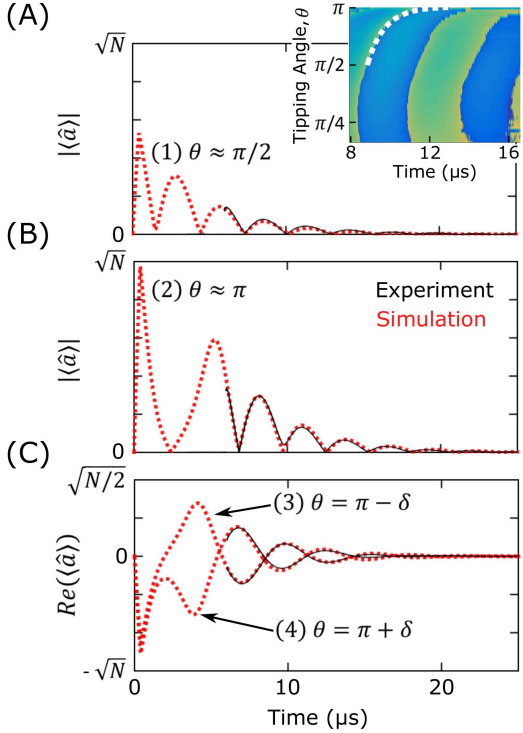}
\captionsetup{justification=raggedright,singlelinecheck=false}
\caption{Amplitude of the pseudospin microwave emission, $\mid \langle \hat{a}\rangle \mid$, during the FID for selected powers (tipping angles) taken from Fig. \ref{fig:FIDmwPowerDep}.  (A) Trace (1) corresponds to a rotation angle of $\theta\approx\pi/2$ (initial pseudospin state $\mid S, 0\rangle$) and trace (2) is for $\theta\approx\pi$ (initial pseudospin state $\mid S, S\rangle$).  All oscillations in trace (1) occur at a rate close to $g\sqrt{N}$.  (B) In contrast, trace (2) evolves more slowly with a long delay in the FID oscillations occurring at the first minimum where the pseudospin reaches full inversion.  The rest of the FID trace oscillates at $g\sqrt{N}$ so that this initial delay shows up as an overall phase shift.  (Inset in the top plot) A zoomed in part of the two-dimensional FID plot from Fig. \ref{fig:FIDmwPowerDep} with a fit (white dashed curve) of the oscillation phase delay according to Eq. \ref{eq:Delay}. (C)~Real part of the pseudospin microwave emission, Re$\left(\langle\hat{a}\rangle\right)$, for the FID traces (3) and (4) from Fig. \ref{fig:FIDmwPowerDep}.  The traces correspond to slight under/over rotation with respect to the fully inverted state of the pseudospin ($\theta\approx\pi\pm\delta$, with $\delta=9$ degrees).  The phases of the microwave emission are opposite in these two traces since the pseudospin precession reverses direction in the under-rotated case but continues in the same direction in the over-rotated case.}
\label{fig:FIDmwPowerDep1DTrace} 
\end{figure}

	The initial z-projection ($M$) of the pseudospin is set by the tipping angle (power) of the microwave pulse.  This tipping angle (defined as polar angle $\theta$ with respect to $\mid S, -S\rangle$ in Fig. \ref{fig:TCCartoon}C) was varied through several full rotations of the pseudospin and the resulting free induction decays (real components) are plotted in Fig. \ref{fig:FIDmwPowerDep} (top) accompanied by Maxwell-Bloch simulations (bottom).  Rotation angles marked as $\pi$, 3$\pi$, and 5$\pi$ on the left in Fig. \ref{fig:FIDmwPowerDep} correspond to the pseudospin being fully inverted (near $\mid S, S\rangle$) after photons from the microwave pulse are fully absorbed by the pseudospin.  We observe that the period of the FID oscillations does not depend on the rotation angle (initial value of $M$) and is always $g\sqrt{N}$.  However there is an overall phase delay in the onset of the oscillations when approaching the fully inverted points at the top of the Bloch sphere (Fig. \ref{fig:FIDmwPowerDep}).  This shift is more clearly seen when comparing the FID traces in Fig. \ref{fig:FIDmwPowerDep1DTrace}A and B for $\theta \approx \pi / 2$ and $\theta \approx \pi$, respectively.

	This delay is explained by considering the pseudospin evolution, starting from $\mid S, M\rangle$, as an avalanche process with the pseudospin transitioning down the ladder of states in the superradiant subspace towards its ground state (Fig. \ref{fig:TCCartoon}C).  Initially there are no photons in the cavity and the emission is spontaneous.  As the pseudospin de-excites, a larger population of photons builds up in the cavity and the pseudospin emission becomes stimulated.  The total time for this process to occur is dominated by the slow time it takes to spontaneously emit the first few photons in the avalanche which is longest near full inversion (Fig. \ref{fig:FIDmwPowerDep1DTrace}B), since the transition probability of the pseudospin is the smallest there.\cite{Dicke1954}  We use the expression derived in \cite{Gross1982301}:

\begin{equation}
\begin{split}
t_D(M) &\approx \frac{2}{\Gamma} \left[ \frac{1}{S-M}+\ldots+\frac{1}{N}\right]\approx \frac{2}{\Gamma}\log{\frac{2S}{S-M}}\\
&=\frac{2}{\Gamma}\log{\frac{2}{(1+\cos{\theta})}},
\end{split}
\label{eq:Delay}
\end{equation}
to fit the delay observed in our experiment (white dashed line in inset of Fig. \ref{fig:FIDmwPowerDep1DTrace}A).  We assume an initial state $\mid S, M \rangle$ with $M \approx S$.  $\Gamma$ is the total spontaneous emission rate from $N$ independent spins.  We extract $\Gamma =1.0$ MHz from this fit.  The longest delay we were able to achieve was $t_D\approx3.5~\mu$s corresponding to an initial Z projection of $M=8.8\cdot10^{12}$ and to the tipping angle $\theta=177^{\circ}$.  The deviation from $180^{\circ}$ was determined by the resolution of the microwave power sweep experiment in Fig. \ref{fig:FIDmwPowerDep}.  

	The oscillation delay is also explained by Eq. \ref{eq:NConcavity}.  With the pseudospin near $M=S$ ($\langle \hat{S}_z\rangle \approx N$) and no photons in the cavity ($n \ll N$), the second term is linear in $N$.  There are also no correlations between spins here ($\langle\hat{S}_+\hat{S}_-\rangle\approx N$) so that the first term is also linear in $N$ giving a slow total evolution ($\ddot{n}\propto g^2 N$) as compared to the other extrema in the FID ($\ddot{n}\propto g^2 N^2$).  

	The phase of the microwave emission during the free induction decay shifts abruptly by $\pi$ as the pseudospin state is varied through the quasi-equilibrium at full inversion around $\theta=\pi$, 3$\pi$, and 5$\pi$ in Fig. \ref{fig:FIDmwPowerDep}.  This phase shift is more clearly seen in Fig. \ref{fig:FIDmwPowerDep1DTrace}C where two traces are shown that correspond to slight underrotation (trace 3) and slight overrotation (trace 4) of the pseudospin relative to full inversion.  This phase shift is explained by looking at the evolution of the pseudospin on the Bloch sphere.  When the microwave drive excites the pseudospin short of the fully inverted state ($\theta=\pi-\delta$, trace 3 in Fig. \ref{fig:FIDmwPowerDep1DTrace}C), just after all of the cavity photons have been absorbed, the pseudospin reverses directions on the Bloch sphere in order to evolve towards its ground state.  The phase of the emitted photons in this case have an opposite phase to the original photons from the microwave drive.  On the other hand, when the cavity photons oversaturate the pseudospin ($\theta=\pi+\delta$, trace 4 in Fig. \ref{fig:FIDmwPowerDep1DTrace}C)  and drive it past the fully inverted state, then the pseudospin will continue to evolve in the same direction as it transitions towards its ground state.  In this case the emitted photons are of the same phase as the initial microwave drive - opposite to the phase of the pseudospin emission in the under-rotated experiment.

The quasi-equilibrium we observe when the pseudospin is at full inversion is a clear indication that the full dynamics of the Maxwell-Bloch equations must be included to model our system.  This is different from the more extensively studied low excitation limit where the pseudospin is always near $M=-S$ and to good approximation $\hat{S}_z$ can be considered to be a constant of motion (Holstein-Primakoff approximation). \cite{1939HolsteinPrimakoff,1989Raizen,1996Childs,Putz2014}  With this approximation the equations of motion are analogous to a system of two coupled linear oscillators and cannot reproduce the behavior we observe near full inversion.  In the low excitation limit, the maxima and minima of the FID both evolve with $\ddot{n}\propto g^2N$ which is much slower than the $\ddot{n}\propto g^2N^2$ behavior we observe here.

%%%%%%%%%%%%%%%%%%%%%\section{\label{sec:Conclusions}Conclusions}%%%%%%%%%%%%%%%%%%%%%%%%%%%%%%%%%%%%%%
In conclusion, we have demonstrated a strongly coupled spin ensemble with uniform spin-cavity coupling and small inhomogeneous broadening allowing the system to be modeled very accurately as a single large pseudospin.  This pseudospin evolves according to the Tavis-Cummings model in the large $N$ limit.  The use of highly enriched $^{28}$Si gives a small inhomogeneous broadening (long T$_2^*$) providing a wide window for viewing the collective pseudospin-cavity dynamics.  In particular we are able to observe the pseudospin-cavity dynamics with a large number of excitations where the Holstein-Primakoff approximation is no longer valid.  We can prepare an arbitrary initial state of the pseudospin $\mid S, M\rangle$ through a combination of optical polarization (determining $S$) and microwave manipulation (determining $M$).  From this we observe the coherent exchange of energy between the pseudospin and the cavity at a rate $g\sqrt{N}$, being independent of M.  Control over M allows us to view, for the first time, the pseudospin dynamics as it approaches a quasi-equilibrium near full inversion.  In particular we observe (and can control) a delay in the pseudospin microwave emission when it is near the quasi-equilibrium point.  We find that this delay scales with the initial state of the pseudospin ($\mid S, M\rangle$) as $\log{\frac{2S}{S-M}}$ which is consistent with an expression derived from the Tavis-Cummings model.  We also observe that the microwave emission has an abrupt $\pi$ phase shift when the pseudospin is at the quasi-equilibrium which is explained by considering the motion of the pseudospin on the Bloch sphere.  These observations offer novel validations of the Tavis-Cummings model.

\section{\label{sec:Acknowledgments} Acknowledgments}
This work was supported by the NSF and EPSRC through the Materials World Network and NSF MRSEC Programs (Grant No. DMR-1107606, EP/I035536/1, and DMR-01420541), and the ARO (Grant No. W911NF-13-1-0179).  MLWT was supported by the Natural Sciences and Engineering Research Council of Canada (NSERC).  The work at keio is supported by KAKENHI (S) No. 26220602 and JSPS Core-to-Core Program.  The authors extend special thanks to Hakan T\"{u}reci and Jonathan Keeling for helpful discussions of the Tavis-Cummings model.

\appendix

\bibliographystyle{prxstyle}
\bibliography{strongcoupling_references_v1_brose} %name of reference file here.

\end{document}